\documentclass[%
 reprint,
 amsmath,amssymb,
 aps,longbibliography
]{revtex4-1}

\usepackage{graphicx}
\usepackage{dcolumn}
\usepackage{bm}
\usepackage{enumitem}

\begin{document}

\preprint{APS/123-QED}

\title{Twin-field quantum key distribution with discrete-phase-randomized sources}
\author{Chun-Mei Zhang$^{1}$}\email{cmz@njupt.edu.cn}
\author{Yi-Wei Xu$^{1}$}
\author{Rong Wang$^2$}
\author{Qin Wang$^{1}$}\email{qinw@njupt.edu.cn}

\affiliation{1 Institute of Quantum Information and Technology, Nanjing University of Posts and Telecommunications, Nanjing, 210003,  China}
\affiliation{2 CAS Key Laboratory of Quantum Information, University of Science and Technology of China, Hefei 230026, China\\}
\date{\today}

\begin{abstract}
Thanks to the single-photon interference at a third untrusted party, the twin-field quantun key distribution (TF-QKD) protocol and its variants can beat the well-known rate-loss bound without quantum repeaters, and related experiments have been implemented recently. Generally, quantum states in these schemes should be randomly switched between the code mode and test mode. To adopt the standard decoy-state method, phases of coherent state sources in the test mode are assumed to be continuously randomized. However, such a crucial assumption cannot be well satisfied in experimental implementations. In this paper, to bridge the gap between theory and practice, we propose a TF-QKD variant with discrete-phase-randomized sources both in the code mode and test mode, and prove its security against collective attacks. Our simulation results indicate that, with only a small number of discrete phases, the performance of discrete-phase-randomized sources can overcome the rate-loss bound and approach that of continuous-phase-randomized sources.
\end{abstract}

\maketitle


\section{Introduction}

Quantum key distribution (QKD) \cite{BB84,E91} can provide two legitimate peers Alice and Bob with information-theoretic secret keys, even in the presence of an eavesdropper Eve. Due to the advantage of theoretic security, a lot of QKD experiments aimed at high rate and long distance have been completed\cite{BB84-Exp-Shields-Optica-2017,QKD-Exp-WS-2GHz,BB84-Exp-421,MDI-Exp-404,MDI-Exp-NP-Pirandola2015}. Despite these impressive achievements, their performance is restricted by the fundamental rate-loss limit\cite{LinearBound,PLOB}, which was believed to be true for any point-to-point QKD without quantum repeaters. Surprisingly, this limit was broken by the revolutionary idea of twin-field QKD (TF-QKD)\cite{TF-Nature}. Subsequently, variant TF-QKD protocols \cite{TF-Tamaki2018,PMQKD-Ma-PRX,TF-SNS-WXB,TF-CCH-PRApp,TF-Curty-npj,PM-LJ-PRA,TF-YHL-SR,TF-WR} were proposed to improve the security, and some of them have been demonstrated experimentally\cite{TF-Exp-MM,TF-Exp-WS,TF-Exp-Lo,TF-Exp-Ma-NP,TF-Exp-LY,TF-Exp-WXB}. At the same time, related theoretic works were extensively studied by researchers to make TF-QKD more applicable to practical channels  \cite{TF-Asymmetric-WXB,TF-Asymmetric-PRA,TF-Asymmetric-NJP,TF-Asymmetric-NJP2020}.

Among all these TF-QKD schemes, quantum states are randomly switched between the code mode and test mode to guarantee the security. The decoy-state method \cite{Decoy-Lo,Decoy-Wang} is adopted in the test mode to estimate the eavesdropper's information or the phase error rate. In the standard decoy-state scheme, phases of a coherent source should be continuously randomized in the range of $[0,2\pi )$, so that the source can be regarded as a classical mixture of photon-number states. In this sense, Eve cannot attack the signal and decoy states with different manners since the signal and decoy states are indistinguishable for her. However, such a crucial requirement cannot be well satisfied in experimental implementations. In practice, the imperfection of phase modulation may be exploited by Eve. In an extreme case, if the phases of a coherent state are not randomized, Eve can launch an unambiguous-state-discrimination (USD) measurement to distinguish between the signal and decoy states, and then attacks them differently without Alice and Bob's awareness \cite{USDAttack-PRA-TYL}. 

Generally, one can turn a laser on and off to generate the expected continuous-phase-randomized pulses, however, the high-speed quantum random number experiments \cite{QRNG-Xu-OE,QRNG-Abell-OE-2014} demonstrated residue correlations existed between phases of adjacent pulses. Hence, the aforementioned doing is doubtful, which should not be adopted in practical QKD systems. An alternative way is to actively modulate the phases of a coherent source with a phase modulator. Intuitively, modulating infinite number of phases can approximate the continuous randomized phases, which nevertheless cannot be realized in practice. Even $1000$ phases modulated in \cite{MDI-Exp-Lo} are finite and discrete, which cannot be directly thought as continuous phases.
%
%

In this paper, to bridge the gap between theory and practice in phase randomization, we propose a TF-QKD protocol with discrete-phase-randomized sources inspired by the idea in \cite{PMQKD-Ma-PRX,TF-WR,MDI-SDP-Lim,BB84-DPR-Cao}, and prove its security against collective attacks. In our protocol, Alice (Bob) prepares coherent states with discrete phases randomly chosen from $\left\{ {\frac{{2\pi x}}{M}|x = 0,1, \cdots ,M - 1} \right\}$ ($\left\{ {\frac{{2\pi y}}{M}|y = 0,1, \cdots ,M - 1} \right\}$) both in the code mode and test mode, where $M$ denotes the number of discrete phases modulated by Alice (Bob). Simulation results indicate that, with only a small number of discrete phases, the performance of our protocol can overcome the rate-loss bound and approach that of continuous-phase-randomized sources, which is very practical and can be realized with current technology. 

\section{TF-QKD with discrete-phase-randomized sources}
First, we give the procedure of TF-QKD with discrete-phase-randomized sources as follows:
\begin{enumerate}[nosep]
\item[$(1)$] Alice (Bob) randomly chooses the code mode or test mode in each trial. 
\item[$1.1)$]If a code mode is selected, Alice (Bob) randomly chooses a key bit $k_a$ ($k_b$) and a random number $x$ ($y$) to prepare a coherent state $\left| {\sqrt \mu  {e^{i({k_a}\pi  + \frac{{2\pi x}}{M})}}} \right\rangle $ ($\left| {\sqrt \mu  {e^{i({k_b}\pi  + \frac{{2\pi y}}{M})}}} \right\rangle $), where ${k_a},{k_b} \in \{ 0,1\} $, $x,y \in \{ 0,1,2, \cdots ,M - 1\} $, $\mu$ denotes the intensity of coherent states, and $M$ denotes the number of discrete phases modulated by Alice (Bob).
\item[$1.2)$]  If a test mode is selected, Alice (Bob)  randomly chooses an intensity $\xi_a$ ($\xi_b$) and a random number $x$ ($y$) to prepare a coherent state $\left| {\sqrt {{\xi _a}} {e^{i\frac{{2\pi x}}{M}}}} \right\rangle $ ($\left| {\sqrt {{\xi _b}} {e^{i\frac{{2\pi y}}{M}}}} \right\rangle $), where ${\xi _a},{\xi _b} \in \{ \mu ,\nu ,\omega \} $.
\item[$(2)$] Alice and Bob transmit their quantum states to a third party Eve. An honest Eve interferes the states on a 50:50 beam splitter, directs the two output pulses to two threshold detectors $L$ and $R$, and announces her measurement results. Only three measurement results are acceptable by Alice and Bob, that is, only detector $L$ clicks, only detector $R$ clicks, and no detectors click. Here, both detectors $L$ and $R$ click is considered as no detectors click. Note that, the security of this protocol does not rely on whether Eve is honest or not.
\item[$(3)$] They repeat the above steps many times. For the successful measurement (only detector $L$ or $R$ clicks), Alice and Bob announce the corresponding mode for each trial. (For simplicity, assuming $M$ is an even number.)
\item[$3.1)$]For trials in the code mode, Alice and Bob announce their $x$ and $y$. If it is the matched trial $x=y$ or opposite trial $x = y \pm \frac{M}{2}$, they keep $k_a$ and $k_b$ as their sifted key bit. If it is the opposite trial $x = y \pm \frac{M}{2}$, Bob flips his key bit $k_b$. Moreover, if Eve announces only detector $R$ clicks, Bob flips his key bit $k_b$. 
\item[$3.2)$] For trials in the test mode, Alice and Bob announce $\xi_a$, $x$, $\xi_b$, and $y$, and they only keep the matched trials $x=y$ or opposite trials $x = y \pm \frac{M}{2}$ with the same intensity ${\xi _a} = {\xi _b}$ to calculate gains. 
\item[$(4)$] Alice and Bob perform key reconciliation and privacy amplification to get the final secret keys.
\end{enumerate}

Based on Devetak-Winter's bound \cite{DWBound}, the final secret key rate of our protocol is
\begin{equation}
R \ge \frac{2}{M}{Q^\mu }(1 - fH({e^\mu }) - I_{AE}^\mu ),
\end{equation}
where $\frac{2}{M}$ is the sifting factor, $Q^\mu$ and $e^\mu$ denote the total gain and error rate of quantum states in the code mode, $H(x) =  - x{\log _2}x - (1 - x){\log _2}(1 - x)$ is the binary Shannon entropy, $f$ denotes the inefficiency of key reconciliation, and $I_{AE}^\mu$ denotes the upper bound of Eve's Holevo information. The detailed security proof is shown in Appendix \ref{SecurityProof}.

\section{Simulation}
For a typical implementation of TF-QKD \cite{TF-Ma-PRApp}, we assume the dark count rate per pulse and the detection efficiency of single photon detectors are $10^{ - 8}$ and $20\% $ respectively, the inefficiency of key reconciliation is $1.1$, and the intrinsic misalignment error is $1.5\% $. With these system parameters, we simulate the performance of TF-QKD with different number of discrete phases. To maximize the performance of our protocol, the intensities of $\mu$ and $\nu$ are optimized, and the intensity of $\omega$ is set to be $0$. 

\begin{figure}[h]
	{\includegraphics[width=\columnwidth]{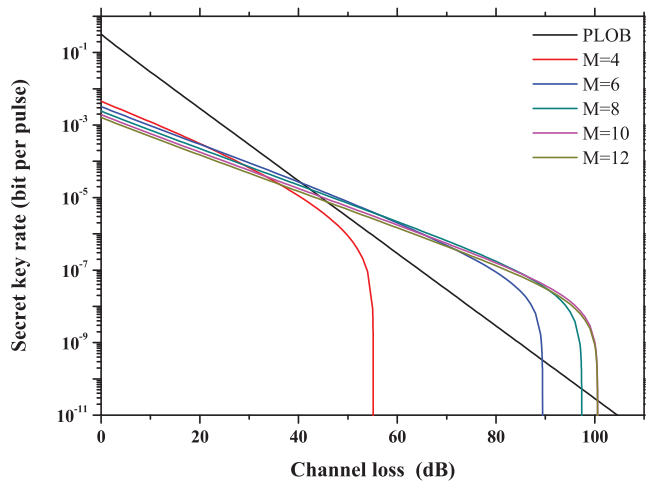}}
	\caption{Results of the secret key rate with respect to the channel loss between Alice and Bob. The black line represents the PLOB bound\cite{PLOB}, and the curves represent the secret key rates of TF-QKD with $M = 4,6,8,10,12$ respectively. Since the secret key rate of $M \to \infty $  tends to $0$, we do not present it here. 
		\label{ThreeDcy}}
\end{figure}

The secret key rate of our protocol is illustrated in Fig. \ref{ThreeDcy}, and the PLOB bound \cite{PLOB} is plotted in comparison, With discrete phase randomization both in the code mode and test mode, the performance of $M=4$ cannot break the PLOB bound, the performance of $M=6$ can break the PLOB bound, and the maximal channel loss of $M=10$ and $M=12$ approaches that of $M \to \infty $. We can also see that, with the increase of $M$, the tolerable channel loss becomes higher due to the relatively accurate estimation of Eve's information, as a tradeoff, the secret key rate becomes lower due to the sifting factor $\frac{2}{M}$. Hence, in practical implementations of TF-QKD, modulating finite discrete phases is adequate to ensure both the security and performance.

\begin{figure}[h]
	{\includegraphics[width=\columnwidth]{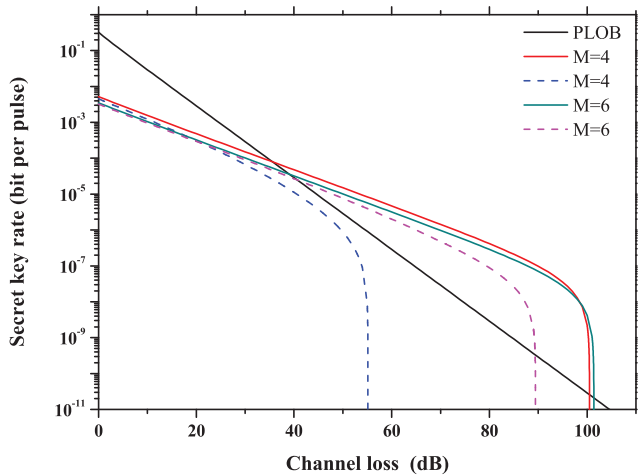}}
	\caption{Comparison results of our protocol and \cite{TF-WR} with $M=4,6$.  The dashed curves denote the results of our protocol, and the solid curves denote the results of \cite{TF-WR}.
		\label{M4and6}}
\end{figure}

\begin{figure}[h]
	{\includegraphics[width=\columnwidth]{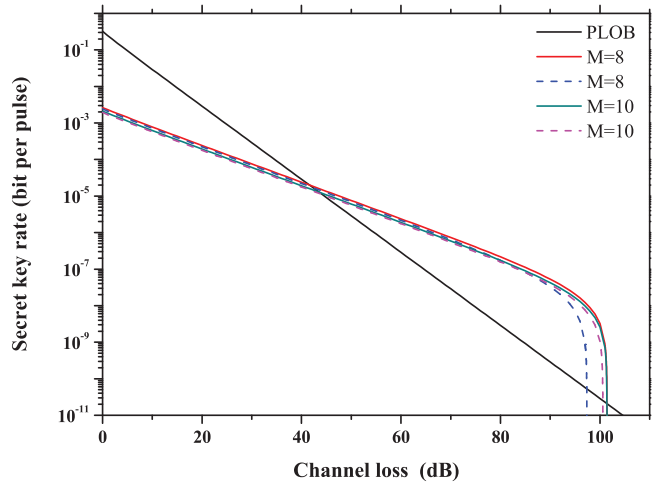}}
	\caption{Comparison results of our protocol and \cite{TF-WR} with $M=8,10$.  The dashed curves denote the results of our protocol, and the solid curves denote the results of \cite{TF-WR}.
		\label{M8and10}}
\end{figure}

We also compare the performance of our protocol and the protocol in \cite{TF-WR}. The difference of these two protocols lies in the test mode. \cite{TF-WR} assumes continuous-phase-randomized coherent states with infinite intensities, which is technically intractable to prepare continuous-randomized phases and infinite intensities. At the same time, since the post-selected phases in \cite{TF-WR} are continuous, the sifting probability of phase postselection tends to $0$, which is very challenging to obtain successful sifting events in practice. In contrast, our protocol modulates discrete phases for three-intensity coherent states, which can be easily realized with current technology. In Figs. \ref{M4and6} and \ref{M8and10}, the dashed curves denote the results of our protocol, and the solid curves denote the results of \cite{TF-WR}. When $M=4,6$, our protocol performs worse than \cite{TF-WR}, while when $M=8,10$, our protocol exhibits performance comparable to \cite{TF-WR}. Hence, considering the  practical feasibility and performance, it is sufficient for TF-QKD to modulate three-intensity coherent states with appropriate number of phases in the test mode.

\section{conclusion}
In conclusion, we have proposed a TF-QKD protocol with discrete-phase-randomized sources both in the code mode and test mode, and proved its security in the asymptotic case. Our simulation results indicate that, modulating only a few number of discrete phases (say $M=10$) in TF-QKD can exhibit performance comparable to that of modulating infinite number of continuous phases,  which is more practical and secure in real-life implemetation of TF-QKD. We expect our work can provide a valuable reference for researchers to design TF-QKD systems. 

\begin{acknowledgments}
This work was supported by the National Key Research and Development Program of China (Grant No. 2018YFA0306400), the National Natural Science Foundation of China (Grant Nos. 61705110, 11774180),  the China Postdoctoral Science Foundation (Grant Nos. 2019T120446, 2018M642281) , the Natural Science Foundation of Jiangsu Province (Grant No. BK20170902), and the Jiangsu Planned Projects for Postdoctoral Research Funds (Grant No. 2018K185C).
\end{acknowledgments}

{\it Note added.} During the preparation of this paper, we find that Guillermo et al. \cite{TF-DPRS} posted a TF-QKD protocol with discrete phase randomization on arXiv. However, the methodologies of security analysis and optimization in this paper are distinct from \cite{TF-DPRS}. Furthermore, \cite{TF-DPRS} modulates different number of phases in the code and test modes, and our protocol modulates the same number of phases in these two modes. Hence, when switching between the code mode and test mode, our protocol only needs to modulate intensities of coherent states. In terms of performance, modulating $4$ phases in \cite{TF-DPRS} can break the PLOB bound, while our protocol cannot; modulating $12$ phases in \cite{TF-DPRS} can get close to the infinite phases, while our protocol needs $10$ phases to get close to the infinite phases.

\appendix

\section{Security proof}
\label{SecurityProof}
First, Let us consider the ideal scenario Alice and Bob prepare perfect continuous-randomized phases in the test mode \cite{TF-WR}, which is equivalent to $M \to \infty $ in our protocol. After the intensity and phase announcement, Alice and Bob keep the post-selected states of matched phases ${\phi _a} = {\phi _b}$ or opposite phases ${\phi _a} = {\phi _b} \pm \pi $ with the same intensity ${\xi _a} = {\xi _b} = \xi $. In the case of the matched/opposite phases and ${\xi _a} = {\xi _b} = \xi $, the ideal states shared between Alice and Bob can be expressed as a combination of $n$-photon state ${{{\left| {n, \pm } \right\rangle }_{AB}}\left\langle {n, \pm } \right|}$, that is, 
\begin{equation}
\begin{array}{l}
\rho _{AB}^{ideal} = \frac{1}{{2\pi }}\int_0^{2\pi } {d \phi _a\left| {\sqrt \xi  {e^{i{\phi _a}}}} \right\rangle \left\langle {\sqrt \xi  {e^{i{\phi _a}}}} \right| \otimes \left| {\sqrt \xi  {e^{i{\phi _b}}}} \right\rangle \left\langle {\sqrt \xi  {e^{i{\phi _b}}}} \right|} \\
= \sum\limits_{n = 0}^\infty  {P(n){{\left| {n, \pm } \right\rangle }_{AB}}\left\langle {n, \pm } \right|} ,
\end{array}
\end{equation}
where $P(n) = \frac{{{e^{ - 2\xi }}{{(2\xi )}^n}}}{{n!}}$ denotes the probability of obtaining the $n$-photon state ${\left| {n, \pm } \right\rangle _{AB}} = \frac{1}{{\sqrt {{2^n}n!} }}{({a^\dag} \pm {b^\dag})^n}{\left| {00} \right\rangle _{AB}}$, ${{{\left| {n, + } \right\rangle }_{AB}}}$  (${{{\left| {n, - } \right\rangle }_{AB}}}$) is corresponding to the state of matched (opposite) phases. Obviously, the ideal $n$-photon state ${{{\left| {n, \pm } \right\rangle }_{AB}}\left\langle {n, \pm } \right|}$ is independent on the intensity $\xi$. From the perspective of Eve, she cannot know an $n$-photon state belongs to which intensity. Therefore, the standard decoy-state method can be directly adopted in \cite{TF-WR}. However, it is technically intractable for Alice and Bob to prepare perfect continuous-randomized phases. At the same time, the sifting probability of phase postselection tends to $0$ when $M \to \infty $, which is challenging to obtain successful sifting events in practice. 

On the contrary, in our protocol Alice and Bob prepare only $M$ discrete phases in the test mode. After the intensity and phase announcement, Alice and Bob keep the states of matched phases $x = y$ or opposite phases $x = y \pm \frac{M}{2} $ with the same intensity  ${\xi _a} = {\xi _b} = \xi $. Specifically, in the case of the matched/opposite phases and ${\xi _a} = {\xi _b} = \xi $, the composite states shared by Alice and Bob can be written as a mixture of approximated $k$-photon state ${\left| {\lambda _k^\xi , \pm } \right\rangle _{AB}}\left\langle {\lambda _k^\xi , \pm } \right|$, that is,
\begin{equation}
\begin{array}{l}
{\rho _{AB}} = \frac{1}{M}\sum\limits_{x = 0}^{M - 1} {{{\left| {\sqrt \xi  {e^{i\frac{{2\pi x}}{M}}}} \right\rangle }_A}\left\langle {\sqrt \xi  {e^{i\frac{{2\pi x}}{M}}}} \right|} \\
\otimes {\left| {\sqrt \xi  {e^{i\frac{{2\pi y}}{M}}}} \right\rangle _B}\left\langle {\sqrt \xi  {e^{i\frac{{2\pi y}}{M}}}} \right|\\
= \sum\limits_{k = 0}^{M - 1} {P_M^\xi } (k){\left| {\lambda _k^\xi , \pm } \right\rangle _{AB}}\left\langle {\lambda _k^\xi , \pm } \right|,
\end{array}
\end{equation}
where $P_M^\xi (k) = \sum\limits_{l = 0}^\infty  {\frac{{{e^{ - 2\xi }}{{(2\xi )}^{lM + k}}}}{{(lM + k)!}}} $ is the proportion of obtaining the approximated $k$-photon state ${\left| {\lambda _k^\xi , \pm } \right\rangle _{AB}} = \frac{{{e^{ - \xi }}}}{{\sqrt {P_M^\xi (k)} }}\sum\limits_{l = 0}^\infty  {\frac{{{{(\sqrt {2\xi } )}^{lM + k}}}}{{\sqrt {(lM + k)!} }}} {\left| {lM + k, \pm } \right\rangle _{AB}}$, and $\left| {\lambda _k^\xi , + } \right\rangle _{AB}$ ($\left| {\lambda _k^\xi , - } \right\rangle _{AB}$) is corresponding to the state of matched (opposite) phases. Note that ${\left| {\lambda _k^\xi , \pm } \right\rangle _{AB}}$ is dependent on the intensity $\xi$ for finite $M$, that is, ${\left| {\lambda _k^{\xi_a} , \pm } \right\rangle _{AB}} \ne {\left| {\lambda _k^{\xi_b} , \pm } \right\rangle _{AB}}$ for different intensities ${\xi_a}$ and ${\xi_b}$. Luckily, ${\left| {\lambda _k^\xi , \pm } \right\rangle _{AB}}$ is close to ${{{\left| {k, \pm } \right\rangle }_{AB}}}$, and we can bound their difference with the method in \cite{BB84-DPR-Cao}.


In view of the security against collective attacks, Eve's attack behavior can be described as an arbitrary unitary operation $U$ on the whole composite systems with her ancillary state ${\left| e \right\rangle _E}$ followed by an arbitrary measurement. Concretely, under the representation of the approximated $k$-photon state $\left| {\lambda _k^\xi , + } \right\rangle _{AB}$, Eve's collective attacks can be expressed as
\begin{equation}
\begin{array}{l}
U{\left| {\lambda _k^\xi , \pm } \right\rangle _{AB}}{\left| e \right\rangle _E} = \sqrt {Y_{k, \pm }^{\xi ,L}} \left| {\gamma _{k, \pm }^{\xi ,L}} \right\rangle \left| L \right\rangle  + \sqrt {Y_{k, \pm }^{\xi ,R}} \left| {\gamma _{k, \pm }^{\xi ,R}} \right\rangle \left| R \right\rangle \\
+ \sqrt {Y_{k, \pm }^{\xi ,N}} \left| {\gamma _{k, \pm }^{\xi ,N}} \right\rangle \left| N \right\rangle ,
\end{array}
\label{collectiveattack}
\end{equation}
where $\left| L \right\rangle $, $\left| R \right\rangle $, and $\left| N \right\rangle $ denote Eve's measurement results; $\left| {\gamma _{k, \pm }^{\xi ,L}} \right\rangle $, $\left| {\gamma _{k, \pm }^{\xi ,R}} \right\rangle $, and $\left| {\gamma _{k, \pm }^{\xi ,N}} \right\rangle $ are arbitrary quantum states corresponding to Eve's measurement results $\left| L \right\rangle $, $\left| R \right\rangle $, and $\left| N \right\rangle $; $Y_{k, \pm }^{\xi ,L}$, $Y_{k, \pm }^{\xi ,R}$ and $Y_{k, \pm }^{\xi ,N}$ satisfying $Y_{k, \pm }^{\xi ,L} + Y_{k, \pm }^{\xi ,R} + Y_{k, \pm }^{\xi ,N} = 1$ are yields corresponding to $\left| L \right\rangle $, $\left| R \right\rangle $, and $\left| N \right\rangle $ given Alice and Bob's composite states ${\left| {\lambda _k^\xi ,\pm} \right\rangle _{AB}}$. Note that Eq.(\ref{collectiveattack}) denotes the most general collective attacks of Eve, which covers the possible attack trying to distinguish between the signal and decoy states, since the unitary operation $U$ and ancillary state ${\left| e \right\rangle _E}$ are arbitrary. Define the total yield of states ${{{\left| {\lambda _k^\xi , + } \right\rangle }_{AB}}\langle \lambda _k^\xi , + |}$ and ${{{\left| {\lambda _k^\xi , - } \right\rangle }_{AB}}\langle \lambda _k^\xi , - |}$ as $Y_k^\xi  = \frac{1}{2}(Y_{k, + }^{\xi ,L} + Y_{k, + }^{\xi ,R} + Y_{k, - }^{\xi ,L} + Y_{k, - }^{\xi ,R})$. Since  ${\left| {\lambda _k^\xi , \pm } \right\rangle _{AB}}$ is dependent on $\xi$, $Y_k^\xi $ is dependent on $\xi$. For different intensities ${\xi _a}$ and ${\xi _b}$, $Y_k^{\xi _a} $ and $Y_k^{\xi _b} $ can be bounded by $\left| {Y_k^{\xi _a} {\rm{ - }}Y_k^{\xi _b} } \right| \le \sqrt {1{\rm{ - }}F_{{\xi _a} {\xi _b} ,k}^2} $, where ${F_{{\xi _a}{\xi _b},k}} = \left| {\left\langle {\lambda _k^{{\xi _a}}, + |\lambda _k^{{\xi _b}}, + } \right\rangle } \right| = \left| {\left\langle {\lambda _k^{{\xi _a}}, - |\lambda _k^{{\xi _b}}, - } \right\rangle } \right| $ \cite{BB84-DPR-Cao}.

Next, we analyze the security of our protocol by considering the matched trials $x=y$ and opposite trials $x = y \pm \frac{M}{2}$ in the code mode separately, and at last take these two cases together to give the final secret key rate of our protocol. For simplicity, in the following context, we omit the subscripts $A$, $B$ and $E$ if there is no confusion.

{\it \textbf{1. Analysis in the matched trials $x=y$ of the code mode.}} In the code mode, the states prepared by Alice and Bob in the matched trials $x=y$ can be expressed as the following four cases: 
if ${k_a} = {k_b} = 0$, 
\begin{equation}
{\left| {\sqrt \mu  {e^{i\frac{{2\pi x}}{M}}}} \right\rangle \left| {\sqrt \mu  {e^{i\frac{{2\pi x}}{M}}}} \right\rangle {\rm{ = }}\sum\limits_{k = 0}^{M - 1} {{e^{i\frac{{2\pi kx}}{M}}}\sqrt {P_M^\mu (k)} } \left| {\lambda _k^\mu , + } \right\rangle };
\end{equation}
if ${k_a} = {k_b} = 1$, 
\begin{equation}
{\left| { - \sqrt \mu  {e^{i\frac{{2\pi x}}{M}}}} \right\rangle \left| { - \sqrt \mu  {e^{i\frac{{2\pi x}}{M}}}} \right\rangle  = \sum\limits_{k = 0}^{M - 1} {{e^{i(\frac{{2\pi x}}{M} + \pi k)}}\sqrt {P_M^\mu (k)} } \left| {\lambda _k^\mu , + } \right\rangle };
\end{equation}
if $k_a=0$ and $k_b=1$,
\begin{equation}
{\left| {\sqrt \mu  {e^{i\frac{{2\pi x}}{M}}}} \right\rangle \left| { - \sqrt \mu  {e^{i\frac{{2\pi x}}{M}}}} \right\rangle  = \sum\limits_{k = 0}^{M - 1} {{e^{i\frac{{2\pi x}}{M}}}\sqrt {P_M^\mu (k)} } \left| {\lambda _k^\mu , - } \right\rangle };
\end{equation}
if $k_a=1$ and $k_b=0$,
\begin{equation}
{\left| { - \sqrt \mu  {e^{i\frac{{2\pi x}}{M}}}} \right\rangle \left| {\sqrt \mu  {e^{i\frac{{2\pi x}}{M}}}} \right\rangle  = \sum\limits_{k = 0}^{M - 1} {{e^{i(\frac{{2\pi x}}{M} + \pi k)}}\sqrt {P_M^\mu (k)} } \left| {\lambda _k^\mu , - } \right\rangle }.
\end{equation}
Correspondingly, considering collective attacks, we have
\begin{equation}
\begin{array}{l}
U{\left| {\sqrt \mu  {e^{i\frac{{2\pi x}}{M}}}} \right\rangle }{\left| {\sqrt \mu  {e^{i\frac{{2\pi x}}{M}}}} \right\rangle }{\left| e \right\rangle }\\
= \sum\limits_{k = 0}^{M - 1} {{e^{i\frac{{2\pi kx}}{M}}}\left( {\left| {\psi _{k, + }^{\mu ,L}} \right\rangle {\left| L \right\rangle } + \left| {\psi _{k, + }^{\mu ,R}} \right\rangle {\left| R \right\rangle } + \left| {\psi _{k, + }^{\mu ,N}} \right\rangle {\left| N \right\rangle }} \right)} ,
\end{array}
\label{U00}
\end{equation}
\begin{equation}
\begin{array}{l}
U{\left| { - \sqrt \mu  {e^{i\frac{{2\pi x}}{M}}}} \right\rangle }{\left| { - \sqrt \mu  {e^{i\frac{{2\pi x}}{M}}}} \right\rangle }{\left| e \right\rangle }\\
= \sum\limits_{k = 0}^{M - 1} {{e^{i(\frac{{2\pi kx}}{M} + \pi k)}}\left( {\left| {\psi _{k, + }^{\mu ,L}} \right\rangle {\left| L \right\rangle } + \left| {\psi _{k, + }^{\mu ,R}} \right\rangle {\left| R \right\rangle } + \left| {\psi _{k, + }^{\mu ,N}} \right\rangle {\left| N \right\rangle }} \right)} ,
\end{array}
\label{U11}
\end{equation}
\begin{equation}
\begin{array}{l}
U\left| {\sqrt \mu  {e^{i\frac{{2\pi kx}}{M}}}} \right\rangle \left| { - \sqrt \mu  {e^{i\frac{{2\pi kx}}{M}}}} \right\rangle \left| e \right\rangle \\
= \sum\limits_{k = 0}^{M - 1} {{e^{i\frac{{2\pi kx}}{M}}}\left( {\left| {\psi _{k, - }^{\mu ,L}} \right\rangle \left| L \right\rangle  + \left| {\psi _{k, - }^{\mu ,R}} \right\rangle \left| R \right\rangle  + \left| {\psi _{k, - }^{\mu ,N}} \right\rangle \left| N \right\rangle } \right)} ,
\end{array}
\label{U01}
\end{equation}
and
\begin{equation}
\begin{array}{l}
U\left| { - \sqrt \mu  {e^{i\frac{{2\pi x}}{M}}}} \right\rangle \left| {\sqrt \mu  {e^{i\frac{{2\pi x}}{M}}}} \right\rangle \left| e \right\rangle \\
= \sum\limits_{k = 0}^{M - 1} {{e^{i(\frac{{2\pi kx}}{M} + \pi k)}}\left( {\left| {\psi _{k, - }^{\mu ,L}} \right\rangle \left| L \right\rangle  + \left| {\psi _{k, - }^{\mu ,R}} \right\rangle \left| R \right\rangle  + \left| {\psi _{k, - }^{\mu ,N}} \right\rangle \left| N \right\rangle } \right)} ,
\end{array}
\label{U10}
\end{equation}
where, for ease of notation, we denote $\left| {\psi _{k, \pm }^{\mu ,L/R/N}} \right\rangle  = \sqrt {P_M^\mu (k)Y_{k, \pm }^{\mu ,L/R/N}} \left| {\gamma _{k, \pm }^{\mu ,L/R/N}} \right\rangle $. To further simplify the expressions of Eqs.(\ref{U00}-\ref{U10}), we define their even part as
\begin{equation}
\left| {\psi _{ex, \pm }^{\mu ,L/R/N}} \right\rangle = \sum\limits_{k = 0}^{M/2 - 1} {{e^{i\frac{{2\pi x}}{M}2k}}\left| {\psi _{2k, \pm }^{\mu ,L/R/N}} \right\rangle },
\end{equation}
and odd part as
\begin{equation}
\left| {\psi _{ox, \pm }^{\mu ,L/R/N}} \right\rangle = \sum\limits_{k = 0}^{M/2 - 1} {{e^{i\frac{{2\pi x}}{M}(2k + 1)}}\left| {\psi _{2k + 1, \pm }^{\mu ,L/R/N}} \right\rangle }.
\end{equation} 
Hence, Eqs.(\ref{U00}-\ref{U10}) can be simply expressed as
\begin{equation}
\begin{array}{l}
{U\left| {\sqrt \mu  {e^{i\frac{{2\pi x}}{M}}}} \right\rangle \left| {\sqrt \mu  {e^{i\frac{{2\pi x}}{M}}}} \right\rangle \left| e \right\rangle }\\
\begin{array}{l}
= \left| {\psi _{ex, + }^{\mu ,L}} \right\rangle \left| L \right\rangle  + \left| {\psi _{ex, + }^{\mu ,R}} \right\rangle \left| R \right\rangle  + \left| {\psi _{ex, + }^{\mu ,N}} \right\rangle \left| N \right\rangle \\
+ \left| {\psi _{ox, + }^{\mu ,L}} \right\rangle \left| L \right\rangle  + \left| {\psi _{ox, + }^{\mu ,R}} \right\rangle \left| R \right\rangle  + \left| {\psi _{ox, + }^{\mu ,N}} \right\rangle \left| N \right\rangle ,
\end{array}
\end{array}
\end{equation}
\begin{equation}
\begin{array}{l}
{U\left| { - \sqrt \mu  {e^{i\frac{{2\pi x}}{M}}}} \right\rangle \left| { - \sqrt \mu  {e^{i\frac{{2\pi x}}{M}}}} \right\rangle \left| e \right\rangle }\\
\begin{array}{l}
= \left| {\psi _{ex, + }^{\mu ,L}} \right\rangle \left| L \right\rangle  + \left| {\psi _{ex, + }^{\mu ,R}} \right\rangle \left| R \right\rangle  + \left| {\psi _{ex, + }^{\mu ,N}} \right\rangle \left| N \right\rangle \\
- \left| {\psi _{ox, + }^{\mu ,L}} \right\rangle \left| L \right\rangle  - \left| {\psi _{ox, + }^{\mu ,R}} \right\rangle \left| R \right\rangle  - \left| {\psi _{ox, + }^{\mu ,N}} \right\rangle \left| N \right\rangle ,
\end{array}
\end{array}
\end{equation}
\begin{equation}
\begin{array}{l}
{U\left| {\sqrt \mu  {e^{i\frac{{2\pi kx}}{M}}}} \right\rangle \left| { - \sqrt \mu  {e^{i\frac{{2\pi kx}}{M}}}} \right\rangle \left| e \right\rangle }\\
\begin{array}{l}
= \left| {\psi _{ex, - }^{\mu ,L}} \right\rangle \left| L \right\rangle  + \left| {\psi _{ex, - }^{\mu ,R}} \right\rangle \left| R \right\rangle  + \left| {\psi _{ex, - }^{\mu ,N}} \right\rangle \left| N \right\rangle \\
+ \left| {\psi _{ox, - }^{\mu ,L}} \right\rangle \left| L \right\rangle  + \left| {\psi _{ox, - }^{\mu ,R}} \right\rangle \left| R \right\rangle  + \left| {\psi _{ox, - }^{\mu ,N}} \right\rangle \left| N \right\rangle ,
\end{array}
\end{array}
\end{equation}
and
\begin{equation}
\begin{array}{l}
{U\left| { - \sqrt \mu  {e^{i\frac{{2\pi x}}{M}}}} \right\rangle \left| {\sqrt \mu  {e^{i\frac{{2\pi x}}{M}}}} \right\rangle \left| e \right\rangle }\\
\begin{array}{l}
= \left| {\psi _{ex, - }^{\mu ,L}} \right\rangle \left| L \right\rangle  + \left| {\psi _{ex, - }^{\mu ,R}} \right\rangle \left| R \right\rangle  + \left| {\psi _{ex, - }^{\mu ,N}} \right\rangle \left| N \right\rangle \\
- \left| {\psi _{ox, - }^{\mu ,L}} \right\rangle \left| L \right\rangle  - \left| {\psi _{ox, - }^{\mu ,R}} \right\rangle \left| R \right\rangle  - \left| {\psi _{ox, - }^{\mu ,N}} \right\rangle \left| N \right\rangle .
\end{array}
\end{array}
\end{equation}
Without loss of generality, we first consider the case when Eve's measurement result is $\left| L \right\rangle $. In this case, Eve's state conditional on Alice and Bob's ancillary qubits $AB$ is
\begin{equation}
\begin{array}{l}
\rho _{ABE,x}^{\mu ,L} = \frac{1}{4}P\{ {\left| {00} \right\rangle _{AB}}\}  \otimes P\{ \left| {\psi _{ex, + }^{\mu ,L}} \right\rangle  + \left| {\psi _{ox, + }^{\mu ,L}} \right\rangle \} \\
+ \frac{1}{4}P\{ {\left| {11} \right\rangle _{AB}}\}  \otimes P\{ \left| {\psi _{ex, + }^{\mu ,L}} \right\rangle  - \left| {\psi _{ox, + }^{\mu ,L}} \right\rangle \} \\
+ \frac{1}{4}P\{ {\left| {01} \right\rangle _{AB}}\}  \otimes P\{ \left| {\psi _{ex, - }^{\mu ,L}} \right\rangle  + \left| {\psi _{ox, - }^{\mu ,L}} \right\rangle \} \\
+ \frac{1}{4}P\{ {\left| {10} \right\rangle _{AB}}\}  \otimes P\{ \left| {\psi _{ex, - }^{\mu ,L}} \right\rangle  - \left| {\psi _{ox, - }^{\mu ,L}} \right\rangle \} ,
\end{array}
\end{equation}
where $P\{ \left|  x  \right\rangle \}  = \left|  x  \right\rangle \left\langle  x  \right|$. 
After tracing Bob's qubit out, Eve's state conditional on Alice's qubit $A$ becomes
\begin{equation}
\begin{array}{l}
\rho _{AE,x}^{\mu ,L} = \frac{1}{4}P\{ {\left| 0 \right\rangle _{A}}\}  \otimes (P\{ \left| {\psi _{ex, + }^{\mu ,L}} \right\rangle  + \left| {\psi _{ox, + }^{\mu ,L}} \right\rangle \} \\
+ P\{ \left| {\psi _{ex, - }^{\mu ,L}} \right\rangle  + \left| {\psi _{ox, - }^{\mu ,L}} \right\rangle \} )\\
+ \frac{1}{4}P\{ {\left| 1 \right\rangle _{A}}\}  \otimes (P\{ \left| {\psi _{ex, + }^{\mu ,L}} \right\rangle  - \left| {\psi _{ox, + }^{\mu ,L}} \right\rangle \} \\
+ P\{ \left| {\psi _{ex, - }^{\mu ,L}} \right\rangle  - \left| {\psi _{ox, - }^{\mu ,L}} \right\rangle \} ).
\end{array}
\end{equation}
And then Eve's state can be written as
\begin{equation}
\begin{array}{l}
\rho _{E,x}^{\mu ,L} = \frac{1}{2}(P\{ \left| {\psi _{ex, + }^{\mu ,L}} \right\rangle \}  + P\{ \left| {\psi _{ox, + }^{\mu ,L}} \right\rangle \} \\
+ P\{ \left| {\psi _{ex, - }^{\mu ,L}} \right\rangle \}  + P\{ \left| {\psi _{ox, - }^{\mu ,L}} \right\rangle \} ).
\end{array}
\end{equation}
Hence, the probability that Alice obtains a sifted key bit when Eve announces $\left| L \right\rangle $ in the matched trials is
\begin{equation}
\begin{array}{l}
Q_{x,m}^{\mu ,L} = \frac{1}{2}(|\left| {\psi _{ex, + }^{\mu ,L}} \right\rangle {|^{\rm{2}}}{\rm{ + |}}\left| {\psi _{ox, + }^{\mu ,L}} \right\rangle {{\rm{|}}^{\rm{2}}}\\
{\rm{ + |}}\left| {\psi _{ex, - }^{\mu ,L}} \right\rangle {{\rm{|}}^{\rm{2}}}{\rm{ + |}}\left| {\psi _{ox, - }^{\mu ,L}} \right\rangle {{\rm{|}}^{\rm{2}}}),
\end{array}
\end{equation}
and the corresponding error rate is 
\begin{equation}
\begin{array}{l}
e_{x,m}^{\mu ,L} = \frac{{|\left| {\psi _{ex, - }^{\mu ,L}} \right\rangle {|^{\rm{2}}}{\rm{ + |}}\left| {\psi _{ox, - }^{\mu ,L}} \right\rangle {{\rm{|}}^{\rm{2}}}}}{{|\left| {\psi _{ex, + }^{\mu ,L}} \right\rangle {|^{\rm{2}}}{\rm{ + |}}\left| {\psi _{ox, + }^{\mu ,L}} \right\rangle {{\rm{|}}^{\rm{2}}}{\rm{ + |}}\left| {\psi _{ex, - }^{\mu ,L}} \right\rangle {{\rm{|}}^{\rm{2}}}{\rm{ + |}}\left| {\psi _{ox, - }^{\mu ,L}} \right\rangle {{\rm{|}}^{\rm{2}}}}}\\
= \frac{{|\left| {\psi _{ex, - }^{\mu ,L}} \right\rangle {|^{\rm{2}}}{\rm{ + |}}\left| {\psi _{ox, - }^{\mu ,L}} \right\rangle {{\rm{|}}^{\rm{2}}}}}{{2Q_x^{\mu ,L}}}.
\end{array}
\end{equation}
Hereafter, we use the subscript $m$ in related variables to denote the matched trials if not specified.

Then, with the strong subadditivity of von Neumann entropy and Jesen's inequality, Eve's Holevo information when she announces $\left| L \right\rangle $ is upper bounded by
\begin{equation}
\begin{array}{l}
I_{AE,x,m}^{\mu ,L} \le (1 - e_{x,m}^{\mu ,L})H(\frac{{|\left| {\psi _{ex, + }^{\mu ,L}} \right\rangle {|^{\rm{2}}}}}{{2(1 - e_{x,m}^{\mu ,L})Q_{x,m}^{\mu ,L}}}) + e_{x,m}^{\mu ,L}H(\frac{{|\left| {\psi _{ex, - }^{\mu ,L}} \right\rangle {|^{\rm{2}}}}}{{2e_{x,m}^{\mu ,L}Q_{x,m}^{\mu ,L}}})\\
\le H(\frac{{|\left| {\psi _{ex, + }^{\mu ,L}} \right\rangle {|^{\rm{2}}} + |\left| {\psi _{ex, - }^{\mu ,L}} \right\rangle {|^{\rm{2}}}}}{{2Q_{x,m}^{\mu ,L}}}),
\end{array}
\end{equation}
where $H(x) =  - x{\log _2}x - (1 - x){\log _2}(1 - x)$ is the binary Shannon entropy. For each matched trial when Eve announces $\left| L \right\rangle $, the secret key rate is 
\begin{equation}
R_{x,m}^{\mu ,L} = Q_{x,m}^{\mu ,L}(1 - fH(e_{x,m}^{\mu ,L}) - I_{AE,x,m}^{\mu ,L}),
\end{equation}
where $f$ is the inefficiency of key reconciliation. Consequently, the average secret key rate in the matched trials $x=y$ of the code mode when Eve announces $\left| L \right\rangle $ is
\begin{equation}
{R_m^L} = \frac{1}{M}\sum\limits_{x = 0}^{M - 1} {Q_{x,m}^{\mu ,L}(1 - fH(e_{x,m}^{\mu ,L}) - I_{AE,x,m}^{\mu ,L})} .
\label{AverageRateL}
\end{equation}
Here, we define the average gain of Alice obtaining a sifted key when Eve announces $\left| L \right\rangle $ as ${Q_m^{\mu ,L}} = \frac{1}{M}\sum\limits_{x = 0}^{M - 1} {Q_{x,m}^{\mu ,L}}$, and the corresponding average error rate as ${e_m^{\mu ,L}} = (\sum\limits_{x = 0}^{M - 1} {Q_{x,m}^{\mu ,L}e_{x,m}^{\mu ,L}} )/(\sum\limits_{x = 0}^{M - 1} {Q_{x,m}^{\mu ,L}} )$. With Jensen's inequality, the second term on the right side of Eq. (\ref{AverageRateL}) can be upper bounded as
\begin{equation}
\frac{1}{M}\sum\limits_{x = 0}^{M - 1} {Q_{x,m}^{\mu ,L}H(e_{x,m}^{\mu ,L})}  \le {Q_m^{\mu ,L}}H({e_m^{\mu ,L}}),
\end{equation}
and the third term can be upper bounded as
\begin{equation}
\begin{array}{l}
\frac{1}{M}\sum\limits_{x = 0}^{M - 1} {Q_{x,m}^{\mu ,L}I_{AE,x,m}^{\mu ,L}} \\
\le Q_m^{\mu ,L}H(\frac{{\sum\limits_{k = 0}^{M/2 - 1} {|\left| {\psi _{2k, + }^{\mu ,L}} \right\rangle {|^2}}  + |\left| {\psi _{2k, - }^{\mu ,L}} \right\rangle {|^2}}}{{2Q_m^{\mu ,L}}}),
\end{array}
\end{equation}
where $I_{AE,m}^{\mu ,L} = H(\frac{{\sum\limits_{k = 0}^{M/2 - 1} {|\left| {\psi _{2k, + }^{\mu ,L}} \right\rangle {|^2}}  + |\left| {\psi _{2k, - }^{\mu ,L}} \right\rangle {|^2}}}{{2{Q_m^{\mu ,L}}}})$ denotes the average upper bound of Eve's Holevo information when she announces $\left| L \right\rangle $ .
Hence, the minimum of $R_m^L$ is 
\begin{equation}
{R_m^L} \ge {Q_m^{\mu ,L}}(1 - fH({e_m^{\mu ,L}}) - I_{AE,m}^{\mu ,L}).
\label{MinRateL}
\end{equation}

Similarly, we can obtain the minimum average secret key rate when Eve announces $\left| R \right\rangle $, which is given by
\begin{equation}
{R_m^R} \ge {Q_m^{\mu ,R}}(1 - fH({e_m^{\mu ,R}}) - I_{AE,m}^{\mu ,R}).
\label{MinRateR}
\end{equation}
The meanings of parameters in Eq.(\ref{MinRateR}) are similar to those in Eq.(\ref{MinRateL}), and we do not repeat them here.

Hence, the total secret key rate in the matched trials $x=y$ of the code mode without sifting factor is 
\begin{equation}
R_m = {R_m^L } + {R_m^R }.
\end{equation}
Define the total gain and error rate in the matched trials as ${Q_m^\mu } = {Q_m^{\mu ,L}} + {Q_m^{\mu ,R}}$ and ${e_m^\mu } = \frac{{{Q_m^{\mu ,L}}{e_m^{\mu ,L}} + {Q_m^{\mu ,R}}{e_m^{\mu ,R}}}}{Q_m^\mu }$. To minimize $R_m$, we maximize the upper bounds of ${Q_m^{\mu ,L}}H({e_m^{\mu ,L}}) + {Q_m^{\mu ,R}}H({e_m^{\mu ,R}})$ and ${Q_m^{\mu ,L}}I_{AE,m}^{\mu ,L} + {Q_m^{\mu ,R}}I_{AE,m}^{\mu ,R}$ with Jesen's inequality, that is,
\begin{equation}
{Q_m^{\mu ,L}}H({e_m^{\mu ,L}}) + {Q_m^{\mu ,R}}H({e_m^{\mu ,R}}) \le {Q_m^\mu }H({e_m^\mu }),
\end{equation}
and
\begin{equation}
\begin{array}{l}
{Q_m^{\mu ,L}}I_{AE,m}^{\mu ,L} + {Q_m^{\mu ,R}}I_{AE,m}^{\mu ,R}\\
\le {Q_m^\mu }H(\frac{{\sum\limits_{k = 0}^{M/2 - 1} {|\left| {\psi _{2k, + }^{\mu ,L}} \right\rangle {|^2} + |\left| {\psi _{2k, - }^{\mu ,L}} \right\rangle {|^2} + |\left| {\psi _{2k, + }^{\mu ,R}} \right\rangle {|^2} + |\left| {\psi _{2k, - }^{\mu ,R}} \right\rangle {|^2}} }}{{2{Q_m^\mu }}})\\
= {Q_m^\mu }H(\frac{{\sum\limits_{k = 0}^{M/2 - 1} {P_M^\mu (2k)Y_{2k}^\mu } }}{{{Q_m^\mu }}}),
\end{array}
\end{equation}
where $I_{AE,m}^\mu  = H(\frac{{\sum\limits_{k = 0}^{M/2 - 1} {P_M^\mu (2k)Y_{2k}^\mu } }}{{{Q_m^\mu }}})$ denotes the upper bound of Eve's Holevo information in the matched trials of the code mode.

As a consequence, the total secret key rate in the matched trials of the code mode with the sifting factor can be minimized as
\begin{equation}
{R_m} \ge \frac{1}{M}{Q_m^\mu }(1 - fH({e_m^\mu }) - I_{AE,m}^\mu ),
\label{RateMatched}
\end{equation}
where $\frac{1}{M}$ denotes the sifting factor of the matched trials.

{\it \textbf{2. Analysis in the opposite trials $x = y \pm \frac{M}{2}$ of the code mode.}} In the code mode, the states prepared by Alice and Bob in the opposite trials $x = y \pm \frac{M}{2}$ can be expressed as the following four cases: 
if ${k_a} = {k_b} = 0$, 
\begin{equation}
{\left| {\sqrt \mu  {e^{i\frac{{2\pi x}}{M}}}} \right\rangle \left| {-\sqrt \mu  {e^{i\frac{{2\pi x}}{M}}}} \right\rangle {\rm{ = }}\sum\limits_{k = 0}^{M - 1} {{e^{i\frac{{2\pi kx}}{M}}}\sqrt {P_M^\mu (k)} } \left| {\lambda _k^\mu , - } \right\rangle };
\end{equation}
if ${k_a} = {k_b} = 1$, 
\begin{equation}
{\left| { - \sqrt \mu  {e^{i\frac{{2\pi x}}{M}}}} \right\rangle \left| {  \sqrt \mu  {e^{i\frac{{2\pi x}}{M}}}} \right\rangle  = \sum\limits_{k = 0}^{M - 1} {{e^{i(\frac{{2\pi x}}{M} + \pi k)}}\sqrt {P_M^\mu (k)} } \left| {\lambda _k^\mu , - } \right\rangle };
\end{equation}
if $k_a=0$ and $k_b=1$,
\begin{equation}
{\left| {\sqrt \mu  {e^{i\frac{{2\pi x}}{M}}}} \right\rangle \left| {  \sqrt \mu  {e^{i\frac{{2\pi x}}{M}}}} \right\rangle  = \sum\limits_{k = 0}^{M - 1} {{e^{i\frac{{2\pi x}}{M}}}\sqrt {P_M^\mu (k)} } \left| {\lambda _k^\mu , + } \right\rangle };
\end{equation}
if $k_a=1$ and $k_b=0$,
\begin{equation}
{\left| { - \sqrt \mu  {e^{i\frac{{2\pi x}}{M}}}} \right\rangle \left| {-\sqrt \mu  {e^{i\frac{{2\pi x}}{M}}}} \right\rangle  = \sum\limits_{k = 0}^{M - 1} {{e^{i(\frac{{2\pi x}}{M} + \pi k)}}\sqrt {P_M^\mu (k)} } \left| {\lambda _k^\mu , + } \right\rangle }.
\end{equation}

Similar to the analysis in the previous matched case $x=y$, the secret key rate in the opposite trials $x = y \pm \frac{M}{2}$ of the code mode can be estimated as
\begin{equation}
{R_o} \ge \frac{1}{M}{Q_o^\mu }(1 - fH({e_o^\mu }) - I_{AE,o}^\mu ),
\label{RateOpposite}
\end{equation}
where the subscript o denotes the opposite trials, and the meanings of other parameters in Eq. (\ref{RateOpposite}) are similar to those in Eq.(\ref{RateMatched}).

{\it \textbf{3. Analysis in both the matched and opposite trials of the code mode.}} 
Combining both the matched and opposite trials, the secret key rate of our protocol is
\begin{equation}
R = {R_m} + {R_o}.
\end{equation}
Define the total gain and error rate of our protocol as ${Q^\mu } = \frac{1}{2}(Q_m^\mu  + Q_o^\mu )$ and ${e^\mu } = \frac{{Q_m^\mu e_m^\mu  + Q_o^\mu e_o^\mu }}{{Q_m^\mu  + Q_o^\mu }}$. To minimize the secret key rate, we maximize the upper bounds of $Q_m^\mu H(e_m^\mu ) + Q_o^\mu H(e_o^\mu )$ and $Q_m^\mu I_{AE,m}^\mu  + Q_o^\mu I_{AE,o}^\mu $ with Jesen's inequality, that is,
\begin{equation}
Q_m^\mu H(e_m^\mu ) + Q_o^\mu H(e_o^\mu ) \le 2{Q^\mu }H({e^\mu }),
\end{equation}
and
\begin{equation}
Q_m^\mu I_{AE,m}^\mu  + Q_o^\mu I_{AE,o}^\mu  \le 2{Q^\mu }H(\frac{{\sum\limits_{k = 0}^{M/2 - 1} {P_M^\mu (2k)Y_{2k}^\mu } }}{{{Q^\mu }}}),
\end{equation}
where $I_{AE}^\mu  = H(\frac{{\sum\limits_{k = 0}^{M/2 - 1} {P_M^\mu (2k)Y_{2k}^\mu } }}{{{Q^\mu }}})$ denotes the total upper bound of Eve's Holevo information. 

Consequently, the final secret key rate of our protocol is
\begin{equation}
R \ge \frac{2}{M}{Q^\mu }(1 - fH({e^\mu }) - I_{AE}^\mu ),
\end{equation}
where $\frac{2}{M}$ denotes the total sifting factor of both matched and opposite trials, and the numerical routine to maximize $I_{AE}^\mu $ is given by
\begin{equation}
\begin{array}{l}
{\max I_{AE}^\mu  = H(\frac{1}{{{Q^\mu }}}\sum\limits_{k = 0}^{M/2 - 1} {P_M^\mu (2k)Y_{2k}^\mu } )}\\
\begin{array}{l}
s.t.\\
{Q^\xi }{\rm{ = }}\sum\limits_{k = 0}^{M - 1} {P_M^\xi (k)Y_k^\xi } ,\xi  \in \{ \mu ,\nu ,\omega \} 
\end{array}\\
{\left| {Y_k^{{\xi _a}}{\rm{ - }}Y_k^{{\xi _b}}} \right| \le \sqrt {1{\rm{ - }}F_{{\xi _a}{\xi _b},k}^2} ,{\xi _a} \ne {\xi _b}}\\
{\sum\limits_{k = 0}^{M/2 - 1} {P_M^\mu (2k)Y_{2k}^\mu }  \le \frac{{{Q^\mu }}}{2}.}
\end{array}
\end{equation}
\\
\bibliography{apssamp}

\end{document}